\documentclass{PoS}

\title{Managing Many Simultaneous Systematic Uncertainties}

\ShortTitle{Managing Many Simultaneous Systematic Uncertainties}

\author{\speaker{Luca Lista}\thanks{The speaker is note PI of the project {\sc Insights}, funded by the European Union's Horizon 2020 research and innovation programme, call H2020-MSCA-ITN-2017, under Grant Agreement n. 765710.}\\
        INFN Sezione di Napoli\\
        E-mail: \email{luca.lista@na.infn.it}}

\author{Agostino De Iorio\\
        Universit\`a degli Studi di Napoli Federico II and INFN Sezione di Napoli\\
        E-mail: \email{deiorio@na.infn.it}}

\author{Alberto Orso Maria Iorio\\
        Universit\`a degli Studi di Napoli Federico II and INFN Sezione di Napoli\\
        E-mail: \email{oiorio@na.infn.it}}

\abstract{Recent statistical evaluations for High-Energy Physics measurements, in particular those at the Large Hadron Collider, require careful evaluation of many sources of systematic uncertainties at the same time. While the fundamental aspects of the statistical treatment are now consolidated, both using a frequentist or a Bayesian approach, the management of many sources of uncertainties and their corresponding nuisance parameters in analyses that combine multiple control regions and decay channels, in practice, may pose challenging implementation issues, that make the analysis infrastructure complex and hard to manage, eventually resulting in simplifications in the treatment of systematics, and in limitations to the result interpretation. Typical cases will be discussed, having in mind the most popular implementation tool, {\sc RooStats}, with possible ideas about improving the management of such cases in future software implementations.}

\FullConference{XIII Quark Confinement and the Hadron Spectrum - Confinement2018\\
		31 July - 6 August 2018\\
		Maynooth University, Ireland}

\begin{document}

\section{Systematics uncertainties and nuisance parameters}

The dependence of a probabilistic model on sources of systematic uncertainty is modeled in terms of nuisance parameters.
Those parameters may be known from external measurements with some uncertainty.
Data samples can constrain nuisance parameters and reduce the original uncertainties.

Different approaches are adopted in Bayesian or frequentist applications, but the resulting effect is similar.

Assume a signal-extraction problem based on a data sample $x$ modeled by
parameter(s) of interest $\mu$ and nuisance parameters $\theta$.
$\mu$ is in many cases the so-called signal strenght, i.e.: the ratio of the measured cross section
and the corresponding theory prediction.

Under the Bayesian approach, the posterior probability density for the unknonw parameters $\mu$ and $\theta$ is~\cite{lista}:
\begin{equation}
  P(\mu,\theta\,|\,x) = \frac{
    L(x;\,\mu, \theta)\pi(\mu,\theta)
  }{\int L(x;\,\mu^\prime, \theta^\prime)\pi(\mu^\prime,\theta^\prime)\,\mathrm{d}\mu^\prime\mathrm{d}\theta^\prime
    }\ .
\label{eq:prob}
\end{equation}
From Eq.~(\ref{eq:prob}), the probability density of the parameter of interest $\mu$ alone is given
by integrating $P(\mu,\theta\,|\,x)$ over the nuisance parameters $\theta$:
\begin{equation}
  P(\mu\,|\,x) = 
\int  P(\mu,\theta\,|\,x)\,\mathrm{d}\theta = \frac{
    \int L(x;\,\mu, \theta)\pi(\mu,\theta)\,\mathrm{d}\theta
  }{\int L(x;\,\mu^\prime, \theta^\prime)\pi(\mu^\prime,\theta^\prime)\,\mathrm{d}\mu^\prime\mathrm{d}\theta^\prime
    }\ .
  \label{eq:prob1}
\end{equation}

Under the frequentist approach, the preferred choice of test statistic is the profile likelihood:
\begin{equation}
  \lambda({\mu}) = \frac{L(\mu,\hat{\hat{\theta}})}{L(\hat{\mu},\hat{\theta})}\ ,
\end{equation}
where $\hat{\mu}$ and $\hat{\theta}$ are the best-fit value of the parameters $\mu$ and $\theta$, respectively,
and $\hat{\hat{\theta}}$ is the best-fit value of $\theta$ for a fixed value of $\mu$,
given the data sample $x$.

The distribution of $q_\mu = -2\ln\lambda(\mu)$, or other variations of this test statistic, are used to
determine the signal strength parameter $\mu$ and/or to set upper limits to the new signal yield.

The distribution of the test statistic for $\mu=0$ may be asymptotically approximated to a $\chi^2$
distribution with one degree of freedom, in the case of a single parameter of interest~\cite{cowan}.
This result is due to Wilks' theorem.

\section{Simultaneous fits}

A complementary dataset, or control sample, $y$, may be used to constrain nuisance parameters $\theta$.
This could be the case of calibration data, background estimates from independent data samples, etc.

Statistical problems can be formulated in terms of both the main data sample ($x$) and the control sample ($y$)
assumed to be statistically independent, with a likelihood function determined as the product of the likelihoods
of the two samples:
\begin{equation}
  L(x, y;\,\mu,\theta) = L_{x}(x;\,\mu,\theta) L_{y}(y;\,\mu,\theta)\ ,
\end{equation}
where $L_y$ does not depend on $\mu$ only if there is no signal contamination in the control sample.

Control samples data are not always available in realistic cases, like
calibrations from test beam, data stored in different formats or  analyzed with different software framework, etc.

A simple case may be modeled with a simplified probability density function (PDF) model, given the `nominal' value $\theta^{\mathrm{nom}}$,
that could be a Gaussian, log-normal, Gamma, etc.
In this case, the likelihood function becomes:
\begin{equation}
  L(x,\theta^{\mathrm{nom}};\,\mu,\theta) = L_x(x;\,\mu,\theta) L_{\theta^\mathrm{nom}}(\theta^{\mathrm{nom}};\,\theta)\ .
\end{equation}

A real-case example of analysis performed by fitting simultaneously control regions and a signal region
is the single-top cross-section
measurement performed by CMS~\cite{singletop} at center-of-mass energy of 8 TeV.
Effectively, background yields measured from background-enriched regions are extrapolated to signal regions
using scale factors predicted from simulation. Events are categorized according to the number of selected
hadronic jets and number of jets identified as b jets, in the aforementioned measurement.

In many cases, an effective way to model nuisance parameters is to provide distributions
modeled as histograms (templates)
obtained from simulations by varying each source of systematic uncertainty by plus or minus
one standard deviation of the corresponding nuisance parameter. Intermediate values (or
outside the $\pm 1\sigma$ range) are obtained with interpolation (or extrapolation) using
either parabolic or piece-wise linear models.

Systematic uncertainties may affect the rate (i.e.: cross section) or shape (i.e.: distribution) of a process, or both.
Examples are luminosity, pile-up modeling in simulation, jet energy scale, b-tagging efficiency, misidentification probability,
lepton selection, reconstruction and trigger efficiencies, as well as uncertainties related
to theory modeling: individual cross section predictions, shape and normalization due to renormalization and factorization scales,
parton distribution function models, parton shower modeling, generator choice, etc.
Uncertainty may also be due to the limited size of Monte Carlo generated simulation samples.

\section{Software implementations}

Most of the methods adopted in High Energy Physics are implemented in the {\sc RooStats} C++ framework
using convenient modeling of PDF via the {\sc RooFit} package~\cite{roofit}, released as part of the {\sc Root} toolkit~\cite{root}.

PDFs from templates are derived from {\sc Root} histograms ({\tt RooHistPdf} class).
Such PDF models, together with data and parameter definitions, are stored in a convenient file format
using the class {\tt Roo\-Work\-space}.

Asymptotic approximations from~\cite{cowan} are available and allow to save CPU time avoiding intensive toy Monte Carlo generation.

Many analyses in the CMS experiment use a command-line, datacard-driven, python-powered tool originally developed for the combination of
multiple Higgs production and decay channels. Code and documentation are open to public access~\cite{cmscombine}.

A datacard language allows to define the analyzed channels and the signal and background processes.
Nuisance parameters are associated via datacards to individual channels and processes, and their PDF models and nominal values
are defined.

Data and simulated distributions are stored as histograms. Special care should be given to naming conventions
that are used to identify histograms related to specific processes, channels, and with the proper
one-sigma up or down variations of nuisance parameters.
Bookkeeping may become an issue for complex cases:
histograms may be arranged in different files with overloaded names, or in the same files with different names or
in the same file but different {\sc Root} sub-directories.
Separators, usually underscores, are used in histogram titles in order to match tags with various meanings.

Limited simulation statistics in each bin is also a source of uncertainty:
one parameter per bin implies many parameters in the model.
Considering only the uncertainty in the bin content of the least-populated bins may speed up
the computation considerably.

In some cases, backgrounds in signal region are constrained from control region
scaled by bin-dependent factors:
\begin{equation}
  h_i^{\mathrm{sig}} = h_i^{\mathrm{bkg}}\,\alpha_i\ ,
  \end{equation}
where the scale factors $\alpha_i$ are determined from Monte Carlo samples.
Histogram content in each bin depends on the value of nuisance parameters.
Scaled histograms can be represented by a customized {\tt RooAbsPdf} object.
The {\sc RooFit} helper class {\tt RooFormulaVar} may help,
with the caveat that formulae are encoded into strings, which may require convoluted
code in complex cases, and bugs in the string definition are only spotted at run time.

In some real case applications, automatic data-cards generation may simplify the problem.
Large data cards can be automatically generated with ad-hoc software that anyway
constitute one extra layer on top of CMS Higgs combine tool.

The organization of parameters into categories may simplify the definition of the problem.
Parameters may be common to groups of distributions, e.g.:
\begin{itemize}
\item Common to all spectra:
  \begin{itemize}
  \item Luminosity, jet-energy scale, b-tag, ...
  \end{itemize}
\item Common to a process:
  \begin{itemize}
  \item Theory uncertainties (renormalization and factorization scale, affect both shape and rate)
  \end{itemize}
\item Common to a decay channel:
  \begin{itemize}
  \item Muon, electron efficiencies (reconstruction, isolation, trigger)
  \end{itemize}
\item Specific to a single spectrum:
  \begin{itemize}
  \item Statistical uncertainty from simulation in each bin
  \end{itemize}
\end{itemize}

More easily management of the most commonly used cases may be
approached with possible extensions of the CMS Higgs combine interface,
that can be potentially promoted as common HEP tool that could eventually
even be released in the {\sc Root} toolkit.

\section{The Insights Project}

{\sc Insights}, International Training Network of Statistics for High Energy Physics and Society~\cite{insights},
is a 4-year Marie Sk\l{}odowska-Curie Innovative Training Networks project for the career development of 12 Early Stage Researchers (ESRs) at 10 partner institutions across Europe.
INSIGHTS is focused on developing and applying latest advances in statistics, and in particular machine learning, to particle physics
CERN is part of the network with deep interconnection with the ROOT development team.

{\sc Insights}' Early-Stage Researchers have been selected and will shortly start working on different statistical tools and applications.
One of the projects proposes development for the presented problem.

\section{Conclusions}

Most of data analyses at the Large Hadron Collider, both precision measurements and search for physics beyond the Standard Model, require simultaneous statistical analysis of many data samples in order to constrain systematic uncertainties.
Managing the achieved complexity requires a substantial amount of coding and challenges the structure of the present software interfaces.
Ad-hoc solutions and mini-framework are implemented in experiment and for specific analyses.
A common implementation in the framework of {\sc RooFit}/{\sc RooStats}/{\sc Root} tools is desirable in order to simplify the management of many applications.

\end{document}